\documentclass[12pt]{article}
\newcommand{\be}{\begin{equation}}
\newcommand{\ee}{\end{equation}}
\newcommand{\kk}{$K\overline{K}$ }
\newcommand{\kpkm}{$K^+K^-$ }
\newcommand{\pp}{$\pi^+\pi^-$ }
\newcommand{\mpp}{$m_{\pi\pi}$ }
\newcommand{\et}{$\eta$ }
\newcommand{\downp}{"down--flat" }
\newcommand{\downs}{"down--steep" }
\newcommand{\upp}{"up--flat" }
\newcommand{\ups}{"up--steep" }
\newcommand{\ro}{$\rho(770)$ }

\title{THEORETICAL CONSTRAINTS ON INTERACTION AMPLITUDES OF LIGHT MESONS}

\author{R. Kami\'nski, L. Le\'sniak\thanks{Presented by L. Le\'sniak 
at the Meson 2000 Conference, Cracow, Poland,
 May 19-23, 2000} ~ and K. Rybicki \\ 
\small {Henryk Niewodnicza\'nski Institute of Nuclear Physics,}
 \\ \small{ PL 31-342 Krak\'ow, Poland}}
 
\begin{document} 
\maketitle

\begin{abstract}

We impose unitarity constraints on the $S$--wave isoscalar $\pi\pi$ amplitudes
extracted from the analysis of the $\pi^- p \rightarrow \pi^+ \pi^- n$ data
which have been measured by the CERN--Cracow--Munich collaboration on a 
transversely polarized target
at 17.2 GeV/c $\pi^-$ momentum. Two "steep" solutions contain a narrow 
$S$--wave $f_{0}(750)$ resonance under the \ro and exhibit a considerable 
inelasticity $\eta$ which is
in disagreement with the four pion production data below the $K\overline{K}$ 
threshold. We impose $\eta\equiv 1$ for all data points and examine four sets
of solutions for the $S$--wave isoscalar phase-shifts. The \downp and \upp
solutions easily pass the $\eta \equiv 1$ constraint but the remaining 
\downs and \ups are eliminated. We conclude that the 17.2 GeV data
cannot be described by a relatively narrow $f_{0}(750)$.

\end{abstract}

  
\vspace{6mm} 

Scalar meson spectroscopy is a subject of many phenomenological analyses in
which a construction of interaction amplitudes between
light pseudoscalar mesons (like \pp, \kpkm and other pairs of mesons) is very
important. The
spectrum of scalars is poorly known \cite{pdg} but an agreement on the existence
 of its 
lowest member $f_0(400-1200)$, also called $\sigma$ meson, is now rather common.
At higher energies there exist isoscalars $f_0(980)$, $f_0(1370)$ and 
$f_0(1500)$ found in various production processes. Nature of scalar mesons is 
naturally related to a spectrum of scalar glueballs since a mixing of the  
$q {\bar q}$ states with gluonia can enrich a number of the observed scalar
resonances \cite{qcd}.

A final success of phenomenological analyses in systematization of the existent 
experimental data depends quite substantially on application of the appropriate 
theoretical constraints on multichannel amplitudes. For example, using relations
 coming from parity or isospin symmetry of strong interactions can lead to an
important reduction of a number of independent scattering amplitudes. In some 
channels like \pp one can apply chiral symmetry constraints and the relations 
following from the crossing symmetry. Analyticity of the coupled channel 
amplitudes is also a very important property. The masses and widths of the
resonances can be essentially obtained in a model-independent way if they are
extracted from positions of the T-matrix poles present in all the relevant decay
and production channels. The dispersion relations
serve as a tool to construct mesonic amplitudes like those appearing in 
Roy's equations of the $\pi\pi$ $S$ and $P$ waves. One should also mention a 
particular role played by constraints following from unitarity 
of the S-matrix. Limitations on the phenomenological amplitudes coming from 
unitarity requirement will be discussed in the analysis presented below.

Let us briefly recall the results of our phenomenological analysis \cite{klr}
of the CERN--Cracow--Munich data \cite{ccm} on the reaction 
$\pi^- p \rightarrow \pi^+ \pi^- n$ 
obtained at 17.2 GeV/c. In this reaction several \pp partial waves ($S$,
$P$, $D$ and $F$) are important. There are significant contributions of three 
scalar resonances
in addition to leading resonances \mbox{\ro,} $f_2(1270)$ and $\rho_3(1690)$. 
Using the same data Svec claimed that a narrow scalar 
resonance $f_{0}(750)$ exists below the \kk threshold \cite{svec}.
In \cite{klr} an energy independent separation of the $S$--wave pseudoscalar and
pseudovector amplitudes has been performed and we have extracted four solutions
of the $\pi\pi$ scalar--isoscalar phase shifts called "down--flat",
"down--steep", "up--flat" and "up--steep". The labels "down" and "up" refer to a
behaviour of the $S$-wave intensity which in the effective \pp mass range 
between 800 MeV and 980 MeV is smaller for the case "up" than for the case 
"down". The other two-fold ambiguity is related to the fact that a sign of the
$S-P$ phase difference can be chosen in two ways, so the "flat" phase 
shifts are smaller and the "steep"
phase shifts are larger than the $P$--phases near the \ro resonance. Thus the 
"steep" solutions could be related to the $f_{0}(750)$ while the "flat" 
solutions to a broad $f_{0}(500)$ postulated in \cite{klm}.

The $S$--wave isospin 0 $\pi\pi$ amplitude can be written as 
\be
a_0= {\eta e^{2i\delta_0} - 1 \over 2i},
                                             \label{a0et}\ee
where $\eta$ is inelasticity and $\delta_0$ is the isoscalar phase shift.
The unitarity constraint leads to inequality  $\eta \leq 1$ which in 
phenomenological analysis can sometimes be violated due to experimental errors.
In \cite{klr} we have, however, eliminated the solution "down-steep" since the
values $\eta$ reached 2 at the \pp effective mass $m_{\pi\pi}$ near 900 MeV. 
However, the \ups 
solution cannot be eliminated in the same way since the values of $\eta$ are 
smaller in that range and their errors are substantial. Nevertheless, a general
behaviour of
\et for the \ups solution is similar to the \downs solution showing a two--bump
character. In contrast to two previous solutions the remaining 
\downp and \upp solutions exhibit a very smooth behaviour of \et very close to 
1.

 In \cite{klr2} we have 
examined in more detail a range of $m_{\pi\pi}$ between 720 MeV and 820 MeV,
where five points of $\eta$ corresponding to the \ups solution systematically
lie below 1. The average value of
inelasticity in that range is $0.67 \pm 0.17$, well below 1. The 
probability to
find accidentally all five points below 1 is small, equal to 0.002. Therefore we
have looked for inelastic reactions in which four pions can be produced 
below the \kk threshold with the same quantum numbers as those of the $\pi\pi$ 
system. The reactions such as the central $4\pi$ production in the high energy
proton--proton collisions, peripheral $4\pi^0$ or $2\pi^+2\pi^-$ production by 
high 
energy pion beams and multipion production in the antiproton annihilation have
been
considered. We have noticed that there were generally only a few $4\pi$ events
below 1 GeV and that no peak was seen in the $4\pi$ effective mass distribution
near \ro. 
Thus a natural assumption in the analysis of the $\pi\pi$ production data below
990 MeV is that the inelasticity $\eta \equiv 1$. With this theoretical
constraint we have made a new analysis of the \pp isoscalar--scalar phase
shifts obtained from the  $\pi^- p \rightarrow \pi^+ \pi^- n$ data at 
17.2 GeV/c. In \cite{klr} we have extracted the S-wave \pp elastic amplitude
$a_S$ which was related to the isoscalar $a_0$ and isotensor amplitude $a_2$ 
in the following way:                    
\be
a_0 = 3a_S - {1\over 2} a_2.\label{a0}
\ee
We have also assumed that the $a_2$ amplitude is fully elastic and the isotensor
phase shifts are known from the analysis of the 
$\pi^+ p \rightarrow \pi^+ \pi^+ n$ data of \cite{ho}. Now in view of 
experimental errors  we have to modify the values of $a_S$ obtained in 
\cite{klr} to fulfill the postulated equality $\eta \equiv 1$. The minimum 
modification is to multiply $a_S$ by a real factor $n$ such that
\be
\eta^2 = \left\vert 1+2ia_0\right\vert^2 =
\left\vert1+2i(3na_S -{1\over 2}a_2)\right\vert^2 \equiv1.  \label{eta2}
\ee 
This is a quadratic equation for $n$ which has to be solved for each value of
$m_{\pi\pi}$. Existence of roots is equivalent to the elastic unitarity condition
satisfied by $a_0$, namely $Im ~a_0 = \left\vert a_0\right\vert^2$.

We have obtained the following numerical results for the
solutions discussed in \cite{klr2}: values of $n$ exist for all twenty \mpp 
points
of the solution \upp and for 19 points (except of the extreme point at 990 MeV)
corresponding to the solution \downp. However for 7 points of the solution \ups
and for 12 points of the solution \downs the elastic unitarity condition cannot
be satisfied. The remaining points of $n$ vary with \mpp forming a parabolic
shape with a maximum at about 800 MeV. In contrast to the "steep" solutions both
"flat" solutions are well fitted by constants very close to unity in the whole
\mpp range between 600 and 1000 MeV. These facts represent a strong 
argument against our accepting the "steep" solutions as good physical solutions.

Let us now discuss common features of the "flat" solutions and the major
differences between them. There is an initial slow growing of phase shifts with
\mpp above 600 MeV, then at the \kk threshold there is a sudden jump up by more 
than $100^0$ and then a further rise particularly steep near 1400 MeV. This
behaviour of phases has been interpreted in terms of three scalar resonances
$f_0(500)$, $f_0(980)$ and $f_0(1400)$. The parameters of these resonances
have been recently determined from the positions of the T-matrix poles in the
complex energy plane using a three coupled channel model \cite{kll}. The major
differences between the \upp and \downp solutions exist for \mpp between 800
MeV and 1000 MeV where the \upp phase shifts are larger than the \downp ones
by a few tens of degrees, the difference reaching about $45^0$. This fact has 
some consequences
since the values of the $f_0(500)$ mass and width differ by about 50 MeV between
two "flat" solutions (see \cite{kll}). Around the \kk threshold both solutions
approach each other but the \upp phases are systematically larger than the 
\downp phases up to about 1300 MeV.

In conclusion, we have demonstrated that theoretical constraints imposed on the 
meson-meson interaction amplitudes are very useful in elimination of some 
ambiguities found in phenomenological analyses of experimental data. Both 
"steep" solutions of the scalar-isoscalar $\pi\pi$ phase shifts show unphysical 
behaviour in contrast to the two "flat" solutions which satisfy well the 
unitarity tests. As a consequence a narrow resonance $f_0(750)$ can be 
eliminated and the broad $f_0(500)$ is confirmed.


\end{document}